\pdfoutput=1
\documentclass[usenatbib]{mn2e}
\usepackage{apjfonts}
\usepackage{amssymb}
\usepackage{amsmath}
\usepackage{ctable}
\usepackage{url}
\usepackage{fixltx2e} %% forces figure order to remain fixed (otherwise figure*'s can move out-of-order)

\newcommand{\be}{\begin{equation}}
\newcommand{\ee}{\end{equation}}

\newcommand{\msun}{M_{\sun}}

\newcommand{\paperone}{Paper {\small I}}
\newcommand{\papertwo}{Paper {\small II}}

\newcommand{\Lbol}{L_{\rm bol}}

\newcommand\plotonesize[2]
 {\centering \leavevmode \includegraphics[width={#2\columnwidth}]{#1}}
\newcommand{\plotsidesize}[2]
 {\centering \leavevmode \includegraphics[width={#2\textwidth}]{#1}}
\newcommand{\acknowledgments}{\begin{small}\section*{Acknowledgments}\end{small}}
\newcommand\altaffilmark[1]{$^{#1}$}
\newcommand\altaffiltext[1]{$^{#1}$}
\title[AGN in Disk Host Galaxies]{Do We Expect Most AGN to Live in Disks?
\vspace{-0.5cm}}

\author[Hopkins et al.]{
\parbox[t]{\textwidth}{ 
Philip F. Hopkins\altaffilmark{1}\thanks{E-mail:phopkins@caltech.edu}, 
Dale D.\ Kocevski\altaffilmark{2}, 
Kevin Bundy\altaffilmark{3}
} 
\vspace*{6pt} \\
\altaffiltext{1}{TAPIR, Mailcode 350-17, California Institute of Technology, Pasadena, CA 91125, USA} \\
\altaffiltext{2}{University of California Observatories/Lick Observatory, and Department of Astronomy and Astrophysics, University of California, Santa Cruz, CA 95064 USA} \\
\altaffiltext{3}{Kavli Institute for the Physics and Mathematics of the Universe, Todai Institutes for Advanced Study, the University of Tokyo, Kashiwa, Japan 277-8583 (Kavli IPMU, WPI)} 
\vspace{-0.6cm}
}

\date{Submitted to MNRAS, July, 2013\vspace{-0.6cm}}
\begin{document}
\maketitle
\label{firstpage}

\begin{abstract}
\vspace{-0.0cm}
Recent observations have indicated that a large fraction of the low to intermediate luminosity AGN population lives in disk-dominated hosts, while the more luminous quasars live in bulge-dominated hosts (that may or may not be major merger remnants), in conflict with some previous model predictions. We therefore build and compare a semi-empirical model for AGN fueling which accounts for both merger and non-merger ``triggering.'' In particular, we show that the ``stochastic accretion'' model -- in which fueling in disk galaxies is essentially a random process arising whenever dense gas clouds reach the nucleus -- provides a good match to the present observations at low/intermediate luminosities. However it falls short of the high-luminosity population. We combine this with models for major merger-induced AGN fueling, which lead to rarer but more luminous events, and predict the resulting abundance of disk-dominated and bulge-dominated AGN host galaxies as a function of luminosity and redshift. We compile and compare observational constraints from $z\sim0-2$. The models and observations generically show a transition from disk to bulge dominance in hosts near the Seyfert-quasar transition, at all redshifts. ``Stochastic'' fueling dominates AGN by number (dominant at low luminosity), and dominates BH growth below the ``knee'' in the present-day BH mass function ($\lesssim 10^{7}\,\msun$). However it accounts for just $\sim10\%$ of BH mass growth at masses $\gtrsim 10^{8}\,\msun$. In total, fueling in disky hosts accounts for $\sim30\%$ of the total AGN luminosity density/BH mass density. The combined model also accurately predicts the AGN luminosity function and clustering/bias as a function of luminosity and redshift; however, we argue that these are not sensitive probes of BH fueling mechanisms.

\end{abstract}

\begin{keywords}
galaxies: formation --- galaxies: evolution --- galaxies: active --- 
star formation: general --- cosmology: theory
\vspace{-1.0cm}
\end{keywords}

\vspace{-1.1cm}
\section{Introduction}
\label{sec:intro}

%The origin of quasars and active galactic nuclei (AGN) is a critical question in understanding the nature and origins of supermassive black holes (BHs), accretion disks, and -- given the observed correlations between BH and galaxy properties and potential energetics of BH ``feedback'' -- massive galaxy formation.
The existence of tight correlations between black hole (BH) mass and properties of the host galaxy spheroid, including spheroid mass/luminosity \citep{KormendyRichstone95,magorrian,kormendy:2011.bh.nodisk.corr}, velocity dispersion \citep{FM00,Gebhardt00}, and binding energy/potential depth \citep{aller:mbh.esph,hopkins:bhfp.obs,feoli:2010.bhfp.2} have fundamental implications for the growth of BHs and -- given the \citet{soltan82} argument which implies that most BH mass was assembled in luminous quasar phases \citep[e.g.][]{salucci:bhmf,yutremaine:bhmf,hopkins:bol.qlf,shankar:bol.qlf} -- corresponding active galactic nuclei (AGN) activity.

Fueling the most luminous quasars at a level required to grow the BH significantly involves channeling an entire typical galaxy's supply of gas ($\gtrsim10^{9}-10^{10}\,M_{\sun}$) into the central few pc, probably requiring $\sim10^{11}\,\msun$ worth of gas in the central $\sim100\,$pc, on a timescale comparable to the galaxy dynamical time. Thus, it is commonly assumed that this necessitates an extreme violent galaxy-wide perturbation such as a major galaxy merger. And indeed, gas-rich galaxy mergers are observed to fuel at least a substantial fraction of bright quasars \citep[see e.g.][and references therein]{guyon:qso.hosts.ir,dasyra:pg.qso.dynamics,silverman:qso.hosts,bennert:qso.hosts,liu:2009.z2.qso.hosts.mergers,veilleux:ir.bright.qso.hosts.merging,letawe:qso.host.asymmetries,koss:2010.swift.dual.agn,koss:2012.swift.dual.agn}. 
%stockton:3c273.interaction,heckman84:qso.mergers,sanders88:quasars,sanders88:warm.ulirgs,stockton87:qso.fuzz,stocktonridgway:qso.merger.id,hutchingsneff:qso.host.imaging,sanders96:ulirgs.mergers,bahcall:qso.hosts,canalizostockton01:postsb.qso.mergers,hutchings:redqso.lowz,guyon:qso.hosts.ir,dasyra:pg.qso.dynamics,bennert:qso.hosts,hopkins:qso.all,RECENT}. 
Such encounters also convert disks into spheroids and further grow the bulge via centrally concentrated gas inflows in a merger-induced starburst \citep{mihos:cusps,hibbard.yun:excess.light,robertson:fp,naab:gas,cox:kinematics,hopkins:cusps.fp,hopkins:cusps.mergers,hopkins:cusps.ell,hopkins:cores}. As argued in \citet{hopkins:bhfp.theory,hopkins:seyfert.limits,snyder:gc.bh.corr.from.bhfp}, this deepens the central potential, so a merger both directly strips gas of angular momentum (providing a BH fuel source) and also increases the binding energy of that material (and bulge mass/velocity dispersion), meaning the BH will grow larger even if strong feedback ``resists'' inflows, before ``catching up'' to the BH-host relations and self-regulating.

\begin{figure}
    \centering
    %\plotside{lir_functions.ps}
    \plotonesize{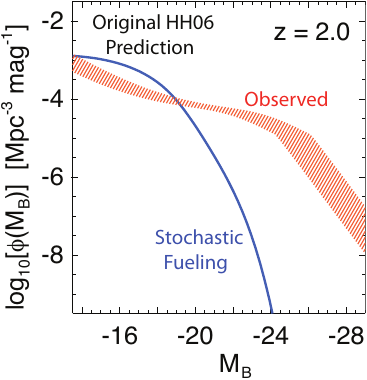}{0.8}
    \caption{Original predicted $z=2$ LF for ``stochastically'' fueled AGN 
    from the models of \paperone, compared to the observed LF at the same 
    redshift fit in \citet{ueda03:qlf}. ``Stochastic fueling'' refers to {\em any} non-major merger triggered accretion of cold 
    gas by AGN (typically in gas-rich disk-dominated galaxies, as opposed to fueling associated with a major merger and substantial bulge growth). 
    The \paperone\ model predicted ``non-major merger'' fueling dominated below luminosities $\Lbol\approx4\times10^{10}\,L_{\sun}$ ($M_{B}\gtrsim-19$). However, \citet{kocevski:2012.candels.agn.hosts} and others (see \S~\ref{sec:intro}) find disk-dominated hosts dominate the population up to at least a factor $\sim10$ higher-$\Lbol$ (close to the ``knee'' in the LF). 
    \label{fig:lf.oldpred}}
\end{figure}

Unfortunately, uniquely identifying observational signatures of ongoing mergers in AGN is incredibly difficult and has been controversial for decades. This is because tidal features are extremely faint and further suppressed by surface-brightness dimming \citep[meaning even the most ``obvious'' mergers are very easily classified as relaxed galaxies; see e.g.][]{lotz:merger.selection,younger:warm.ulirg.evol,puech:2012.disk.survival,snyder:2013.midir.diagnostics.obscured.qso}, mergers are rare so control samples and good statistics are difficult, and the models themselves (almost without exception) predict that the gas inflow rates into the nucleus and subsequent AGN duty cycle peak in the {\em post-merger} phases where the galaxy can easily look like a ``relaxed'' bulge down to optical surface brightnesses $\mu\gtrsim 30\,{\rm mag\,arcsec^{-2}}$ \citep[see][]{dimatteo:msigma,hopkins:zoom.sims,hopkins:sb.agn.delay,hopkins:agn.alignment,li:radiative.transfer,johansson:bh.scalings.in.remergers,snyder:2011.ka.gal.sims}.

Nevertheless, recent observations of AGN host morphologies and colors have suggested that major mergers probably do not fuel most low and intermediate-luminosity AGN, as a large fraction appear in ``normal'' disks \citep{gabor:2009.agn.host.morph,schawinski:2011.agn.in.disks,cisternas:2011.cosmos.agn.nomerger,kocevski:2012.candels.agn.hosts,rosario:2011.candels.agn.colors,civano:2012.agn.host.morph,santini:2012.agn.sf.corr.break,mullaney:2012.secular.bh.from.sf.indicators,treister:2012}. This should perhaps not be surprising. Unlike a bright quasar, fueling a Seyfert (bolometric $L<10^{12}\,L_{\sun}$ or $4\times10^{45}\,{\rm erg\,s^{-1}}$) for a typical $\sim10^{7}$\,yr episode \citep[see][]{martini04} requires a gas supply within the range of just a single or a few giant molecular clouds (GMCs). There are many alternative mechanisms that could sufficiently disturb the gas in the central regions of the galaxy to as to produce such an event. These include minor mergers \citep{hernquist.mihos:minor.mergers,woods:tidal.triggering,woods:minor.mergers,younger:minor.mergers}, secular angular momentum loss in bar/spiral arms \citep[for a review, see][]{jogee:review} or Toomre-unstable ``clumpy'' disks \citep{bournaud:2011.agn.fueling.by.clumps}, steady-state accretion of diffuse (low-density) hot gas \citep[see][and references therein]{allen:jet.bondi.power,best:radio.loudness}, or multi-body interactions with nearby star clusters or other clouds \citep[e.g.][]{genzel:gal.center.review}. All of these processes do occur in galaxies, and should at least indirectly contribute to AGN fueling insofar as they help remove angular momentum from dense gas. 

Many models for the rates and luminosity functions of these processes have been proposed (see references above); however, as far as the central BH is concerned, they are all degenerate in the sense that none {\em directly} interacts with the BH. They instead all serve to drive gas into the galactic nucleus, whereupon some other mechanisms (including torus-scale gas+stellar disk processes and the ``traditional'' AGN accretion disk) must reduce the angular momentum of the gas by an {\em additional} six orders of magnitude before it can be accreted. This complicates any model for galactic-scale ``fueling'' considerably, as it is difficult to imagine any surviving one-to-one correlation between the current BH activity and the galactic state. 

Therefore, \citet{hopkins:seyferts} (hereafter \paperone) attempted to synthesize these processes into a general ``stochastic accretion'' model; rather than modeling every galaxy-scale event in a fully a priori manner (which involves large uncertainties), it is sufficient to know empirically their important effect for ultimate BH fueling, namely the (resulting) distribution of dense gas and its velocity dispersions in the central regions of the galaxy. Individual ``episodes,'' corresponding to the gravitational capture of dense gas (e.g.\ molecular clouds) by the BH directly, occur stochastically but with calculable statistical properties. Coupled to a simple model for AGN feedback, the total duty cycle of AGN as a function of luminosity from these ``non-major merger'' fueling modes can be estimated. \paperone\ argued that this can predict accurately many observed properties of $z\approx0$ Seyferts, including their  host galaxies, luminosity functions (LFs), and duty cycles.

One consequence of such models is the idea (discussed in detail in \citealt{hopkins:seyfert.limits,draper:2012.bh.synthesis.secvsmgr,santini:2012.agn.sf.corr.break,treister:2012}) that there is some characteristic host bulge/BH mass (and corresponding quasar luminosity) below which these more ubiquitous mechanisms dominate AGN fueling (being more common and requiring less bulge growth to deepen the central potential in this mass regime). Above this division, less violent mechanisms are simply inefficient (they may still happen, but they do not sufficiently raise the bulge mass, so BHs quickly self-regulate and do not experience any significant lifetime of high-Eddington ratio growth) and the population requires more extreme mechanisms such as major mergers to build the most massive bulges and (corresponding) BHs. 

Coupling these models to empirical estimates of the evolution of galaxy mass functions, gas fractions, and other quantities, \paperone\ attempted to extend the model predictions to high redshifts. The predicted LF from that paper at $z=2$ is shown in Fig.~\ref{fig:lf.oldpred}. Qualitatively, we see the transition discussed above, with the stochastic mode dominant at low luminosities. 

But the recent observations discussed above find that disk-dominated hosts (i.e.\ candidates for the ``stochastic'' mode, as opposed to post-major merger systems which may not, on average, appear as disks)\footnote{It is important to note that even major galaxy mergers can and do leave disk-dominated remnants under the right circumstances (when they are sufficiently gas rich and have favorable initial orbital parameters; see \citealt{springel:spiral.in.merger,robertson:disk.formation,hopkins:disk.survival}). However, if major mergers were the dominant AGN fueling mechanism, any plausible distribution of orbital parameters (combined with the gas fractions estimated observationally in these populations) would at least produce a significant enhancement of bulge-dominated or ``bulge-enhanced'' galaxies relative to a control population at the same stellar mass \citep[see][]{hopkins:disk.survival.cosmo}. This is not observed except at higher AGN luminosities, as we will discuss further in the text.} dominate the population even at luminosities an order-of-magnitude larger than the ``transition point'' predicted in Fig.~\ref{fig:lf.oldpred}. 

Clearly, there is something wrong with these models.  However, the \paperone\ model remains a good description of some observations at $z=0$, and captures many of the key processes from simulations which appear to be robust even as resolution and the treatment of AGN, star formation, feedback, and ISM physics has improved \citep[see the comparisons in][]{debuhr:momentum.feedback,debuhr:2012.bal.plus.radpressure,johansson:bh.scalings.in.remergers,choi:2012.bh.fb.idealized}. We therefore, in this paper, re-visit these models for AGN fueling, but attempt to incorporate them into a modern, and observationally-constrained ``population synthesis'' model. This allows us to use more accurate assumptions and models for the evolution of the galaxy population with redshift (including galaxy mass/luminosity functions, merger rates, and gas fraction distributions), to define the ``background'' on which AGN fueling occurs. We also attempt to compile a range of observational constraints of the AGN host galaxy population, spanning redshifts $z\approx0-2$, to develop the most rigorous constraints to date and so construct a better estimate of the integrated contribution of major merger vs.\ non-major merger mechanisms towards BH growth.

\begin{figure*}
    \centering
    %\plotside{lir_functions.ps}
    \plotsidesize{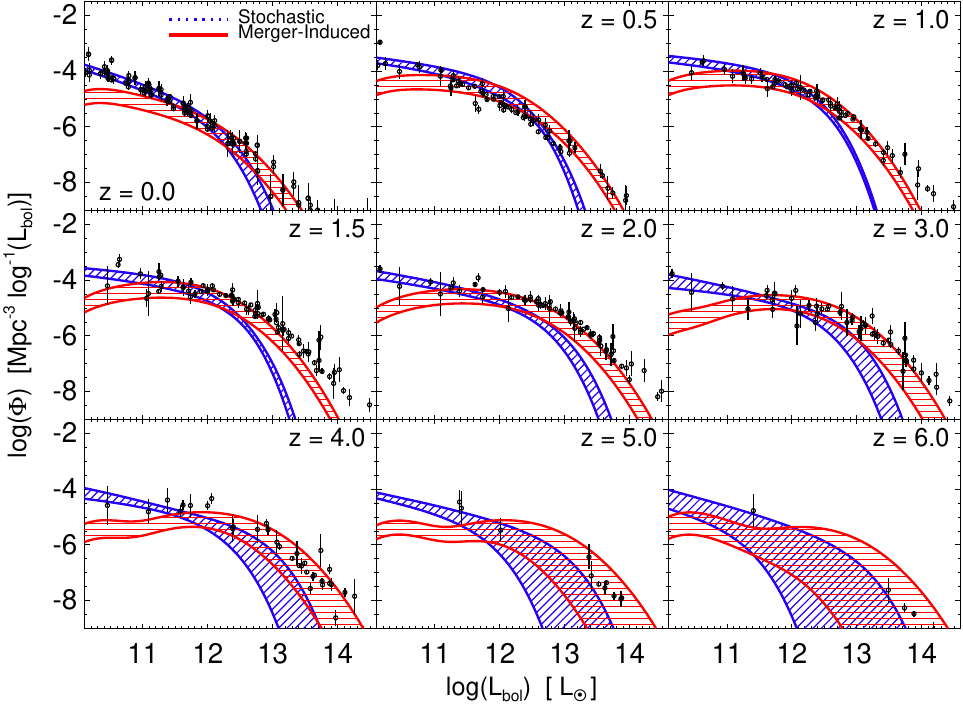}{1.0}
    \caption{Bolometric AGN luminosity functions as a function of redshift. 
    We show the predicted LF of major merger-induced AGN from \papertwo\ (red), and 
    LF of non-major merger ``stochastically'' fueled systems from \paperone\ (blue), with 
    updated observational inputs (stellar mass functions and gas fractions used to construct the 
    model) matching those from \papertwo.
    Shaded ranges reflect the 
    uncertainty from different stellar mass function observations used in constructing the model.
    Black points show the compilation of observational data used to derive 
    bolometric AGN luminosity functions in \citet{hopkins:bol.qlf}.
    \label{fig:lfs}}
\end{figure*}

\begin{figure}
    \centering
    %\plotside{lir_functions.ps}
    \plotonesize{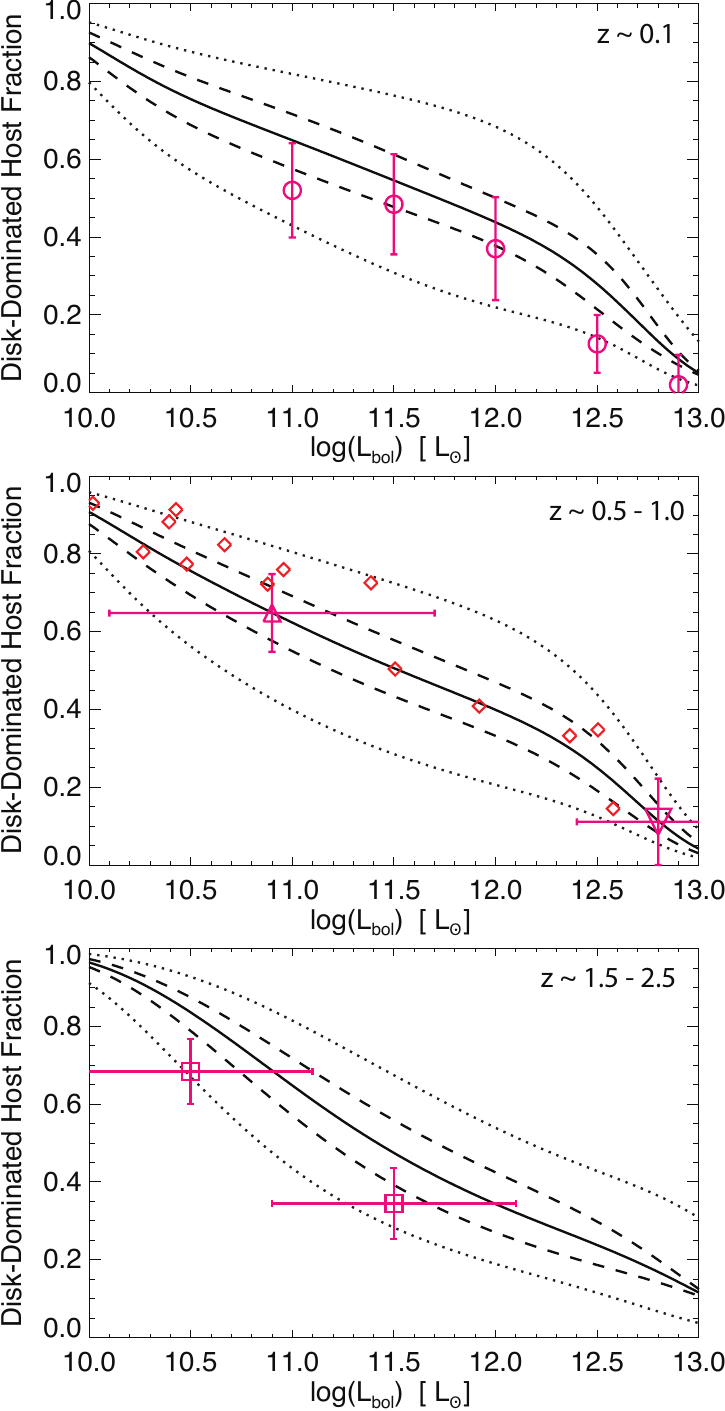}{1.0}
    \caption{Predicted fraction of AGN in the stochastic mode, as a function of luminosity 
    and redshift, from the models in Fig.~\ref{fig:lfs}. The mean of the model range is shown as the solid line with the $\pm1\,\sigma$ (dashed) and ($\pm2\,\sigma$) range (dotted).
    Since the duty cycle of ``stochastically'' fueled systems is dominated by gas-rich disks, 
    and most major merger-triggered systems that induce strong bulge and BH growth will appear as spheroids, 
    we compare the observed fraction of disk-dominated AGN host galaxies in different luminosity/redshift intervals. At low redshift, measured from the PG quasar sample of \citet[][circles]{dunlop:qso.hosts,floyd:qso.hosts}. At intermediate redshift, measured in low-luminosity AGN in COSMOS in \citet[][triangle]{gabor:2009.agn.host.morph,cisternas:2011.cosmos.agn.nomerger}, and in true (Type 2) quasars in \citet[][inverted triangle]{zakamska:qso.hosts,zakamska:mir.seds.type2.qso.transition.at.special.lum,liu:2009.z2.qso.hosts.mergers}. At $z\sim2$, measured in CANDELS AGN host galaxies in the low and high-luminosity sub-samples from \citet[][squares]{kocevski:2012.candels.agn.hosts}. \citet{treister:2012} independently considered a (partially-overlapping) compilation which agrees well with ours (red diamonds in center show $1-f_{m}$, where $f_{m}$ is the fraction of AGN they specifically identify as merger-induced).
    \label{fig:bd}}
\end{figure}

\vspace{-0.5cm}
\section{The Models}
\label{sec:models}

The model we will present here supposes two independent AGN fueling populations. A ``major merger-induced'' population, and a ``stochastic'' population (which essentially includes all non major merger-induced events). We will make the same consistent assumptions about the background population and AGN behavior in fueling events in both cases, but treat the total AGN luminosity function as simply the sum of the predicted duty cycles from both sub-populations.

\vspace{-0.2cm}
\subsection{Merger-Induced Fueling}
\label{sec:merger.fueling.model}

The major merger-induced quasar fueling model here is taken directly from a 
series of papers:  \citet{hopkins:disk.survival.cosmo,hopkins:merger.rates,hopkins:merger.rates.methods,hopkins:ir.lfs}, and \citet{younger:ir.background}. We use the most recent update to the model, 
presented in \citet{hopkins:ir.lfs} (hereafter \papertwo). 
There are three basic components of the model, for which all details are given in \papertwo. Since we are only taking the results from that paper, we will only briefly summarize the key model elements here. 

(1) At a given redshift, we begin with the {\em observed} galaxy mass functions and gas fraction 
distributions. This defines the empirical ``background population'' onto which we will add assumptions for AGN fueling. Of course, other types of models such as semi-analytic models and cosmological simulations attempt to predict these properties {\em a priori}, then further add assumptions about AGN fueling \citep[see, e.g.][and references therein]{
dimatteo:cosmo.bhs,somerville:new.sam,fanidakis:2011.bh.spin.sam}. But this adds considerable uncertainty. Since our focus here is on the AGN population alone, we prefer the \papertwo\ ``semi-empirical'' model approach, which allows us to isolate the assumptions relevant to the AGN population. The actual mass function data are compiled from a range of sources.\footnote{Mass functions measurements are compiled from \citet{bell:mfs,arnouts:2007.type.sep.mfs.to.z2,ilbert:cosmos.morph.mfs,perezgonzalez:mf.compilation,fontana:highz.mfs,marchesini:highz.stellar.mfs,kajisawa:stellar.mf.to.z3}. Where different measurements overlap at the same redshift, we use the differences between them (added with the appropriate quoted error bars) to define the empirical uncertainty in the MF. The compilation is chosen such that there are always at least two overlapping measurements at each redshift. We then interpolate log-linearly between the median MF measurements at each redshift. We combine these with measurements of the mean and scatter in gas fractions as a function of stellar mass and redshift, from \citet{belldejong:disk.sfh,mcgaugh:tf,calura:sdss.gas.fracs,shapley:z1.abundances,erb:lbg.gasmasses,puech:tf.evol,mannucci:z3.gal.gfs.tf,cresci:dynamics.highz.disks,forsterschreiber:z2.sf.gal.spectroscopy,erb:outflow.inflow.masses}. For details, see \papertwo.}

(2) Using a simple abundance-matching halo-occupation model (i.e.\ forcing the population to match observed number densities and clustering; see \citealt{conroy:monotonic.hod}) each observed member of the galaxy population is assigned to a halo, from which the merger rate can be calculated from fits to the cosmological halo-halo merger rates. In other words, from a cosmological simulation, all halo-halo mergers at a specific redshift of interest are identified.\footnote{The specific results here use the halo mergers from the Millenium simulation in \citealt{fakhouri:halo.merger.rates}. However in \citet{hopkins:merger.rates.methods} we compare this to a wide variety of other simulations with varied numerical methods, cosmological parameters, and post-processing method for halo and merger identification; we use this to define a ``theoretical uncertainty'' in the halo merger rate. In the model here, this is added in quadrature to the ``empirical uncertainty'' in the number density of galaxies, to define the total uncertainty in the final merger rate. These uncertainties and comparison of the predicted rates to observations are in \citet{hopkins:merger.rates.methods}.} Each halo is then assigned a galaxy via abundance matching (and a dynamical friction time is assigned between the halo-halo and galaxy-galaxy merger). This leads to the galaxy-galaxy merger rates. Extensive discussion and tests of this methodology are presented in \citet{hopkins:merger.rates.methods}; we simply note here that taking the merger rate {\em directly} from observations gives a  similar result, but with large uncertainties \citep[comparisons with observations and semi-analytic models are in][]{stewart:merger.rates,jogee:merger.density.08,lotz:2011.mgr.rate.comparison}.

(3) For each such ``semi-empirically'' assigned merger, we then attach an AGN fueling model. Specifically, in a series of papers, \citet{hopkins:qso.all,hopkins:merger.lfs,hopkins:bhfp.obs,hopkins:bhfp.theory} use a simple model for AGN accretion rates and feedback to fit the resulting AGN lightcurves in galaxy-galaxy merger simulations as a function of galaxy mass, redshift, and gas fraction of the progenitors. Since we have this information in our mock population, we can then simply assign the corresponding fitted (bolometric) lightcurve (or equivalently, probability of being seen at a given luminosity) to every merger. The exact functional parameterization is given in \papertwo; this is compared to observations \citep[from][]{yu:mdot.dist,kauffmann:new.mdot.dist} of the duty cycle distribution and alternative ``synthesis models'' \citep[from][]{merloni:synthesis.model,shankar:bol.qlf} in \citet{hopkins:mdot.dist}. For our purposes here, the important conclusion in that paper is that the results are all similar, so (within the relatively large uncertainties) it makes relatively little difference which parameterization we adopt. 

We stress that we are not presenting any modifications or revisions to this model; we take the predicted ``merger-induced'' AGN luminosity functions exactly as calculated in \papertwo. Readers interested in how variations within that model affect the results presented here should see \papertwo, Appendix~B.

\vspace{-0.5cm}
\subsection{Stochastic Fueling}

\paperone\ argued that AGN can and should also be triggered stochastically in non-merging systems via a variety of detailed mechanisms. We therefore crudely assign all ``non-major merger'' processes to the ``stochastic fueling'' category. In \paperone, however, this is ``synthesized'' into an estimate of the resulting luminosity function using very crude assumptions about the galaxy population and its redshift evolution. The methodology described above for the merger-induced population provides a much more well-motivated ``background'' onto which we apply the models from \paperone.

The two basic steps are as follows:

(1) At a given redshift, we again begin with the observed galaxy mass functions and gas fraction distributions from \papertwo, identical to the first step in \S~\ref{sec:merger.fueling.model}.

(2) With this information, we apply the model from \paperone\ for the cumulative duty cycle of activity owing to non-major merger fueling mechanisms. This is the major model addition in this paper, to the model presented in \papertwo. 

We begin by assigning a BH mass to every galaxy in the model, at each redshift, according to the simple approximate observed relation: 
\be
M_{\rm BH} \approx 0.0014\,(1+z)^{0.5}\,f_{\rm bulge}\,M_{\ast}
\ee
with a lognormal intrinsic scatter of $\approx0.3\,$dex in $M_{\rm BH}$. This is a purely empirical estimate of a best-fit to a range of observations (\citealt{mclure.dunlop:mbh,peng:magorrian.evolution,woo06:lowz.msigma.evolution,adelbergersteidel:magorrian.evolution,shields06:msigma.evolution,salviander:midz.msigma.evol,treu:msigma.evol,bennert:msigma.evol}; for a recent review see \citealt{kormendy:2013.mbh.mgal.review}). We stress that the relation and scatter are well-anchored at $z=0$, but increasingly uncertain at high redshifts. But theoretical models give similar redshift evolution, mostly owing to the more gas-rich, compact nature of high-redshift hosts \citep[see][]{hopkins:bhfp.theory,johansson:bh.scalings.in.remergers,choi:2012.bh.fb.idealized}. In any case, the results of varying the assumed redshift evolution are shown in \papertwo\ (Figs.~B1 \&\ B2); since it appears in almost identical form in the merger model, it will shift the normalization of both stochastic and merger-triggered AGNs in luminosity $L_{\rm bol}$, but not much alter their {\em relative} behavior, which is most interesting here. The scatter is observed, but has little effect -- it {\em is} important for the abundance of the most massive BHs (above the ``break'' in the galaxy mass function, corresponding to luminosities well above the turnover in the LF; see \papertwo\ Fig.~B3), but we will show that the stochastic mode is sub-dominant in this regime in any case (so assuming any scatter $\lesssim1\,$dex makes little difference). Finally, $f_{\rm bulge}$ is estimated from our galaxy mass functions, but is formally degenerate with the normalization and redshift evolution of the relation; where (at high redshifts) it is poorly determined we simply assume $f_{\rm bulge}\approx0.3$, since this appears to give a good fit to observations of the relation between BH and total stellar mass (see references above).

With BH masses assigned, we need to assign luminosities. Since the triggering mode is ``random'' (on cosmological timescales), it is sufficient to simply assign a duty cycle (probability of observing a given luminosity). This is calculated for the stochastic mode in \paperone, assuming a triggering rate determined by capture of cold gas in the nucleus and subsequent regulation of accretion via feedback. It is shown there that this can be simply parameterized as: 
\be
\label{eqn:PlogL}
\frac{{\rm d}P}{{\rm d}\log{L}} = \alpha\,{\Bigl(}\frac{f_{\rm gas}}{0.1} {\Bigr)}\,{\Bigl(}\frac{L}{L_{\rm Edd}(M_{\rm BH})} {\Bigr)}^{-\beta}
\ee
with $\alpha\approx0.003$ and $\beta\approx0.6$ \citep[see][as well as \paperone]{yu:mdot.dist,shankar:bhmf.integral.model,kelley:broadline.bhmf.methodology,hickox:multiwavelength.agn,bonoli:modeling.qso.clustering.vs.lifetimes.model,kauffmann:new.mdot.dist,trump:2009.type1.agn.mdot.limits}. Here, the parameter $\alpha$ physically represents the duty cycle at high accretion rates, given by the rate at which a collection of Jeans-length sized clouds (on isotropic, virial equilibrium orbits) in a galactic nucleus would intersect the black hole (then multiplied by the time required to accrete the captured mass, i.e.\ $\alpha \sim n_{\rm clouds}\,\pi\,R_{\rm cloud}^{2}\,\sigma_{v,\,{\rm cloud}}\,(\epsilon_{r}\,M_{\rm cloud}\,c^{2}/L)$). The parameter $\beta$ represents the relative amount of time at each accretion rate -- for a simple power-law lightcurve this corresponds to $L\propto t^{-1/\beta}$, with the value here representing the typical behavior of this ``decay'' in each accretion event in simulations. We truncate Eq.~\ref{eqn:PlogL} at $L>L_{\rm Edd}(M_{\rm BH}) \approx 3.3\times10^{4}\,(L_{\sun}/M_{\sun})\,M_{\rm BH}$. Note that in \citet{hopkins:mdot.dist}, this is compared to an extensive ensemble of observational constraints and measurements of the Eddington ratio distribution at $z\approx0-1$, and shown to agree well (especially for moderate luminosity AGN), with relatively little allowed range in $\alpha$ or $\beta$ relative to the theoretically predicted values. Therefore, if we simply adopted a best-fit to the observed $L/L_{\rm Edd}$ distribution at  $z=0$, we would obtain a nearly identical prediction.\footnote{For simplicity, in the plots in this paper we use exactly the formula given in Eq.~\ref{eqn:PlogL}; in \paperone\ a more complicated convolution is presented, to which this is a simplified fitting function. Repeating all of our calculations with the more detailed formulation makes a completely negligible difference to the predictions. Also we should formally introduce a lower limit to Eq.~\ref{eqn:PlogL}, which corresponds to the luminosity above which the duty cycle integrates to unity. However this occurs at such low luminosities that it is irrelevant to any of our calculations.} This duty cycle is simply convolved with the BH mass function to obtain the stochastic-mode LF.

We emphasize that the AGN-centric equations in step (2) were developed in \paperone. What distinguishes our predictions here from those therein is the model for the galaxy population. In \paperone, some very simple assumptions -- many of which appear to be inaccurate in light of observations in the last several years -- were made to extrapolate the model from $z=0$. Implicitly, these would (for the same AGN fueling and feedback model) correspond to a very different distribution of galaxy masses and gas fractions, from that which we develop here. The most important differences are: (1) observations of high-redshift galaxies indicate they are more gas-rich than assumed in \paperone, with gas fractions approaching $\sim50\%$ even in high-mass systems \citep[see e.g.][]{tacconi:high.molecular.gf.highz}; (2) high-mass galaxies are also more abundant at high redshift than was assumed in \paperone, indeed many cosmological simulations and semi-analytic models still under-predict the number density of galaxies with stellar masses $\gtrsim10^{11}\,\msun$ at $z\gtrsim2$ \citep[see][]{hayward:2012.smg.counts}. There was also no explicit model for the merger-induced population in \paperone; here we include that developed in \papertwo.

\begin{figure}
    \centering
    %\plotside{lir_functions.ps}
    \plotonesize{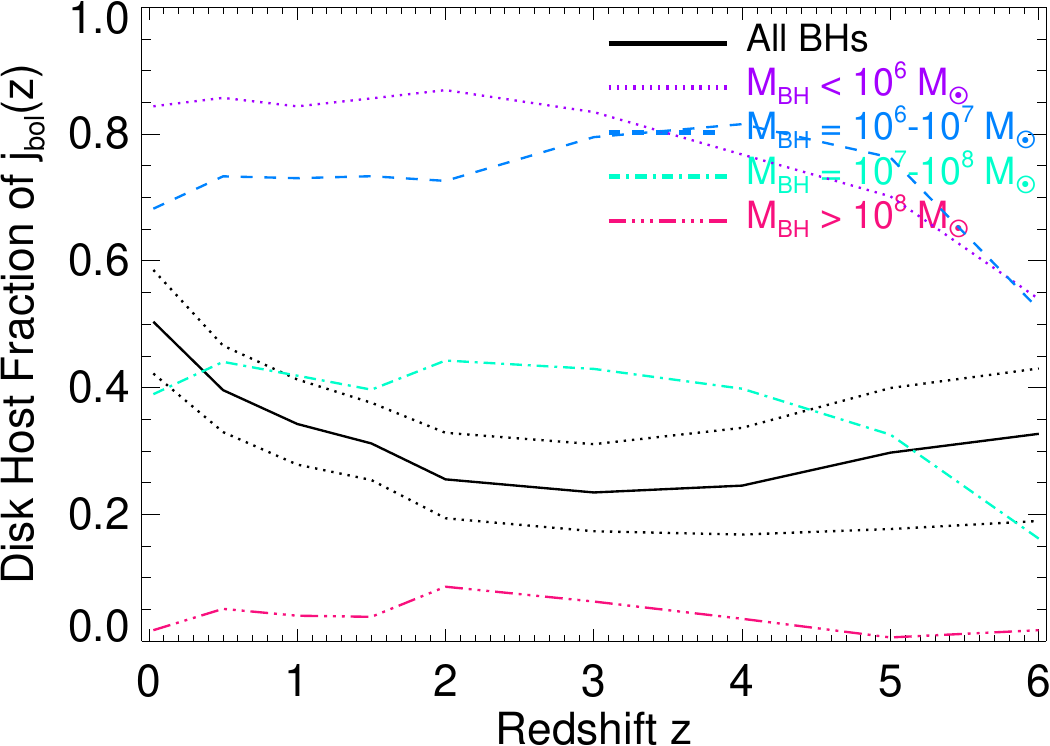}{1.0}
    \caption{Fractional contribution of ``stochastic'' fueling to the AGN luminosity density/integrated BH accretion as a function of redshift. We show the total contribution (solid black), with the uncertainties (within the context of the model) from Fig.~\ref{fig:bd} (dotted black); we also compare the contribution to BH growth in different intervals in BH mass at each redshift (colored, as labeled). The model here is the best-fit to both the bolometric LF and the observed disk/spheroid fractions at each $L$, $z$. The best-fit model predicts $\sim30\%$ of the luminosity density from non-major merger fueling modes (increasing at the lowest redshifts). The non-merger modes completely dominate at low BH masses $\ll 10^{7}\,\msun$, while merger modes dominate at high BH masses $\gg 10^{7}\,\msun$. 
    \label{fig:lumden}}
\end{figure}

\vspace{-0.5cm}
\section{Results}
\label{sec:results}

Fig.~\ref{fig:lfs} plots the predicted AGN luminosity functions from the models for both major merger-induced and ``stochastic'' (non-major merger) mechanisms, at several redshifts. As discussed in \S~\ref{sec:models}, the empirical uncertainties in the galaxy number density, gas fractions, and merger rates at each redshift are added in quadrature to give the ``total uncertainty'' in the model predictions (shaded range in the plots). This should be thought of as the uncertainty owing {\em not} to differences in the AGN fueling models (which might be quite large), but owing to un-related uncertainties in the background galaxy population.

{Of course, there are significant uncertainties in the AGN models themselves. However -- within the context of the models we use here -- these actually contribute surprisingly little to the ``total'' prediction uncertainties. Consider the merger model: a major merger converts an order-unity fraction of the galaxy stellar mass into a bulge, and in any model we consider the final BH must lie near the appropriate $M_{\rm BH}(M_{\rm bulge})$ relation (at that redshift). Since the fueling timescale is short (inflow rates are large), but we cap accretion at the Eddington limit, the contribution to the luminosity function from a merger is dominated by the last Salpeter time near the Eddington limit for $M_{\rm BH}(M_{\rm bulge})$. Thus, if the Eddington limit applies, the uncertainty in the ``merger mode'' duty cycle is dominated by the merger rate, and the uncertainty in the ``merger mode'' characteristic luminosity by the background population (mass, bulge-to-disk and $M_{\rm BH}(M_{\rm bulge})$ distributions). For explicit demonstrations of this we refer to \citet{hopkins:lifetimes.interp,hopkins:qso.all,merloni:synthesis.model,shankar:bhmf.integral.model,bonoli:modeling.qso.clustering.vs.lifetimes.model,somerville:2009.msigma.scatter.qlf}. Of course we could, in principle, assume post-merger systems lie far off the relation for some reason, but we will not consider such models here. For the ``stochastic mode,'' the assumed duty cycle ($\alpha$ in Eq.~\ref{eqn:PlogL}) simply enters linearly in the predicted AGN LF normalization. Theoretically, it is highly uncertain, but empirically, there is not much room for it to vary without violating the constraint that we must match the abundance of low-luminosity AGN. Changing the prediction at high-$L$ without violating this constraint would require a {\em qualitative} change in the model, namely invoking a duty cycle which increases substantially (by a factor of $\sim100$ over the dynamic range plotted) in more massive, bulge-dominated, and gas-poor galaxies -- the opposite of what is seen in simulations and numerical models of secular evolution.} That is not to say there are no uncertainties, just that the uncertainties in the parameters of these models do not have a large effect on our conclusions; we discuss more radical alternative models below.

To avoid uncertainties owing to obscuration, we compare the model predictions (which are really for the bolometric BH accretion rates and luminosities) with empirical estimates of the bolometric (obscuration and wavelength-corrected) AGN luminosity function presented in \citet{hopkins:bol.qlf}.\footnote{This is based on a compilation of observations at wavelengths from the IR through optical, soft and hard X-rays \citep[see e.g.][]{ueda03:qlf,hunt04:z3.faint.qlf,barger:qlf.big,richards05:2slaq.qlf,hasinger05:qlf,lafranca:hx.qlf,brown06:ir.qlf,richards:dr3.qlf,siana:z3.swire.qlf}. An alternative but similar bolometric compilation is presented in \citet{shankar:bol.qlf}, and bolometric LFs from hard X-ray LFs with appropriate corrections are in \citet{aird:hx.qlf,yencho:hx.qlf}; the differences are generally smaller than the model uncertainties in Fig.~\ref{fig:lfs}. Additional observations have been developed since these papers; however they generally overlap with the plotted points except at the highest redshifts ($z\gtrsim5$) where they extend the dynamic range significantly \citep[see][]{mcgreer:2012.z5.qlf}. However at these redshifts the newer data lie well within the (very large) model uncertainties.} 

The sum of the stochastically fueled AGN and merger-induced AGN LFs agrees very well with the observed bolometric LF at most redshifts. This is reassuring, and it also suggests that some large {\em additional} fueling mechanism or driver is not needed to explain the observed demographics.\footnote{At $z\sim1-3$, the total LF at the very highest luminosities $\Lbol\gtrsim10^{14}\,L_{\sun}$ does appear to fall short of the bolometric LF estimates. This is discussed in \papertwo\ (since the predictions are dominated by the merger-induced contribution at these luminosities). It is certainly worth considering that this owes to a deficiency in the AGN fueling/lightcurve models. However, we caution against reading too much into the discrepancy. These are extreme populations with number densities of just a few per cubic Gpc, and so systematic uncertainties in e.g.\ the relevant bolometric corrections and contributions from lensing are very large.} 

Clearly, stochastically fueled systems are predicted to dominate at the lowest luminosities, while merger-induced populations dominate at the highest luminosity. The transition between them occurs at a broadly similar luminosity $\sim 10^{12}\,L_{\sun}$ (the traditional Seyfert-quasar divide) at all redshifts. 

It is important to stress that we have not adjusted or ``fine-tuned'' any parameters in the model here to reproduce the observations. Moreover the stochastic and merger-induced models are independent predictions, so it is encouraging that they appear to accurately sum to reproduce the total luminosity function. However, we should emphasize that the model presented here is not unique, and a combination of many observations is needed to fully break degeneracies in models. The AGN LF alone is a relatively poor constraint on fueling mechanisms: by allowing very different AGN lightcurves, or including minor mergers (and assuming they produce a long duty cycle of low-luminosity activity), it is possible to fit the low-luminosity LF with {\em only} merger-induced fueling \citep[see the models in][]{hopkins:lifetimes.interp,hopkins:merger.lfs,somerville:2009.msigma.scatter.qlf}. On the other hand, by assuming a much stronger ``secular'' mode (in which traditional disk bar instabilities are assumed to channel $100\%$ of the galaxy gas into the nucleus in a single burst -- essentially mimicking a major merger), \citet{fanidakis:2011.bh.spin.sam} show they can plausibly reproduce the high-luminosity LF. And at high redshifts and high-$\Lbol$, we see that the ``allowed range'' owing to uncertainties in galaxy number densities and merger rates is very large -- this means that sufficient degeneracies exist such that the bight, high-redshift LF has little power to constrain fueling models.

In Fig.~\ref{fig:bd}, we use this result to estimate the distribution of host population ``type'' versus mass. Specifically, we plot, at each redshift, the fraction of the population at each $\Lbol$ that are predicted to be fueled in the ``stochastic'' mode (as opposed to the major-merger mode). At all redshifts, we see a continuous increase in the predicted merger-relic AGN population with luminosity, with the merger-mode being negligible at Seyfert luminosities but becoming dominant at QSO luminosities $\gg 10^{12}\,L_{\sun}$. There is some quantitative increase in prevalence of mergers at intermediate luminosities at high redshifts, but the effect is small.

Very crudely, most ``stochastically fueled'' systems should be disk-dominated. To lowest order, this is simply a reflection of the background galaxy population (which, at lower masses where the fueling mode is dominant, is mostly disk-dominated). At second-order, at fixed mass, in the model we adopt (\S~\ref{sec:models}) AGN activity does require a gas supply, so fueling is enhanced in gas-rich systems, which are overwhelmingly disk-dominated (though of course there will be some, albeit rarer, gas-rich spheroids). In contrast, most major merger-fueled systems should be bulge-dominated, since such mergers tend to build large bulges.\footnote{As noted in \S~\ref{sec:intro}, we stress that a sizeable fraction of major mergers will produce disk-dominated galaxies, especially at high-redshifts where the disks are more gas-rich  \citep[see e.g.][]{springel:spiral.in.merger,robertson:disk.formation,hopkins:disk.survival,hopkins:disk.heating,hopkins:stellar.fb.mergers,governato:disk.rebuilding}. However, disk survival in mergers is most efficient at low galaxy masses, where the disks are actually gas-dominated; the large BH masses where the merger-induced mode is dominant imply bulge masses $\gtrsim10^{11}\,\msun$. Moreover large surviving disks generically require conditions (gas distributions and orbits) that {\em suppress} strong inflows, the opposite of the regime we are interested in here where strong bulge growth and AGN fueling will result. As a result, this can be critical for the abundance of disks at low masses \citep{hopkins:disk.survival.cosmo,somerville:new.sam,stewart:disk.survival.vs.mergerrates,puech:2012.disk.survival}, but is probably not the dominant process in the mergers that produce bright quasar activity, of interest here.}

We therefore compare the predicted ``stochastically fueled'' fraction of AGN with the fraction of disk-dominated AGN hosts, as a function of luminosity and redshift. At low redshifts, we compare with the PG quasar sample of \citet{dunlop:qso.hosts,floyd:qso.hosts} (we plot the fraction with best-fit bulge-to-total mass ratio $B/T<0.5$ in bins of $L_{\rm bol}$, estimated from the observed nuclear $V$-band luminosities, with Poisson errors). At $z\sim0.5-1$, we compare the low-luminosity sample from COSMOS studied in \citet{gabor:2009.agn.host.morph} and \citet{cisternas:2011.cosmos.agn.nomerger}; we plot the ``final'' quoted fraction of disk-dominated galaxies in the sample (with the approximate $\sim 10\%$ systematic difference between classifiers quoted therein) and $90\%$ range of $L_{\rm bol}$ estimated from the hard X-ray luminosities. We also compare the sample of true quasars in \citet{zakamska:qso.hosts,zakamska:mir.seds.type2.qso.transition.at.special.lum,liu:2009.z2.qso.hosts.mergers} at $z\sim0.3-0.7$. These are Type-II (obscured) objects whose host morphologies can be determined, of which $1/9$ is a disk galaxy, and the remainder are clearly spheroid-dominated and/or visible late-stage mergers. At $z\sim2$, we compare with the CANDELS sample from \citet{kocevski:2012.candels.agn.hosts}, again using the quoted distribution of visual classifications for their low and high-luminosity samples ($L_{\rm bol}$ estimated here from the hard X-ray luminosities). Note that \citet{treister:2012} consider a similar compilation (partially overlapping the sources we have compiled here); the results from their compilation (using different data sets, bolometric corrections, and classification criteria) agree extremely well with what we plot in Fig.~\ref{fig:bd} at each redshift.

This is only a very rough comparison, to see whether the predictions are at all reasonable given present observational constraints on AGN host galaxy morphologies. Of course, as discussed in \citet{trump:2011.host.agn.morph.discussion}, considerable care is needed regarding the different selection in these samples. We have attempted to match in luminosity and redshift, but other aspects (color, AGN selection criteria, morphological classification method, imaging wavelength) must be investigated in more detail in future work before any rigorous, quantitative ``best-fit'' to these observations can be presented. 

Fig.~\ref{fig:lumden} plots the fractional contribution from the ``stochastic'' mode (predicted from the model), integrated over the luminosity function, to BH growth in different mass intervals and different redshifts.

\begin{figure}
    \centering
    \plotonesize{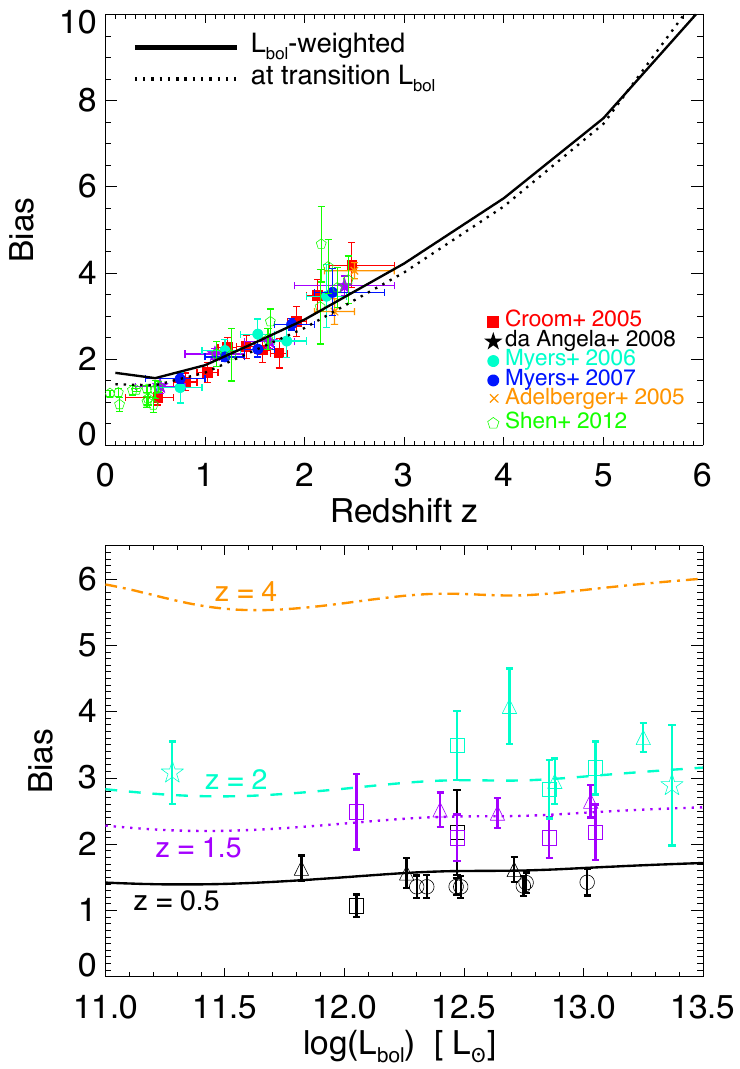}{1.0}
    \caption{Predicted clustering amplitude (linear bias) of AGN populations from the model here. {\em Top:} Mean bias as a function of redshift. We plot the luminosity-density weighted bias (integrated over the LF; solid), and the mean bias at the ``transition'' luminosity where the contribution from stochastic and merger fueling is equal (dotted). We compare to compiled observations of quasar clustering from \citet{hopkins:clustering} and \citet{shen:2012.qso.gal.crosscorr}.$^{\ref{foot:clustering.refs}}$ The two agree well, with bias similar to $\sim1-4\times10^{12}\,\msun$ halos at each redshift. {\em Bottom:} Mean bias as a function of luminosity, at fixed redshifts (specific values shown). We compare observations in narrow luminosity intervals at approximately the same redshifts (denoted by the same colors), from \citet[][circles]{shen:2012.qso.gal.crosscorr}, \citet[][squares]{daangela:clustering}, \citet[][triangles]{myers:clustering}, and \citet[][stars]{adelbergersteidel:lifetimes}. The clustering amplitude predicted is a very weak function of luminosity at all redshifts, in agreement with the observations. In particular, there is no feature or trend marking the ``transition'' in Fig.~\ref{fig:bd}.
    \label{fig:clustering}}
\end{figure}

In Fig.~\ref{fig:clustering}, we use the models to predict the clustering amplitude of AGN populations as a function of redshift and luminosity. Recall that, in the model, every mock AGN has a known host galaxy stellar mass and (via abundance matching) assigned host halo mass. We can then simply adopt the expression for the clustering amplitude (bias) as a function of halo mass and redshift from \citet{shethtormen}, and use this to calculate the mean bias of the population in bins of AGN luminosity and redshift.\footnote{For the clustering calculation, we adopt the WMAP5 cosmological parameters. However within reasonable uncertainties this only has a small systematic effect on the normalization of the bias in Fig.~\ref{fig:clustering}.} We show the mean bias as a function of redshift, for AGN in different luminosity intervals, and the bias as a function of luminosity at specific redshifts. We compare this to the compilation of observations in \citep{hopkins:clustering} and \citet{shen:2012.qso.gal.crosscorr}.\footnote{\label{foot:clustering.refs}\citet{hopkins:clustering} compile the observations from \citet{croom:clustering,adelbergersteidel:lifetimes,myers:clustering.old,myers:clustering,porciani:clustering,daangela:clustering}. \citet{shen:2012.qso.gal.crosscorr} compile the results from \citet{shen:sdss.qso.clustering.update,hickox:multiwavelength.agn,
cappelluti:2010.agn.clustering.local,hickox:2011.qso.clustering.bootes,white:2012.qso.clustering.boss,krumpe:2012.rosat.agn.clustering}. The measurements of clustering amplitude as a function of luminosity at fixed redshift are compiled from \citet[][circles at $z\approx0.5$]{shen:2012.qso.gal.crosscorr}, \citet[][squares at $z\approx0.5,\,1.5,\,2.0$]{daangela:clustering}, \citet[][triangles at $z\approx0.5,\,1.5,\,2.0$]{myers:clustering}, and \citet[][stars at $z\approx2.0$]{adelbergersteidel:lifetimes}.}

The agreement with observations is good. Unfortunately, it appears that clustering is not a strong constraint on models of AGN fueling mechanisms. For example, the trend of bias with redshift, either for AGN near the ``knee'' in the LF or weighted across the LF, is similar in models which assume only merger fueling \citep{lidz:clustering,hopkins:clustering}, only secular (non-merger) fueling (\citealt{fanidakis:2011.bh.spin.sam}, Croton et al., in preparation), or which make no statement about fueling but only assume a random duty cycle independent of galaxy properties \citep{croton:2009.qso.dm.sham,conroy:2013.qso.toy.model}. And we see here that the predicted ``transition'' between the stochastic mode and merger mode does not imprint any characteristic feature in the clustering as a function of luminosity (at a given redshift). 

Finally, we should note (as discussed in \papertwo), that since the synthesis model here essentially {\em assumes} the BH-host correlations observed in order to ``populate'' systems (and the simulations to which the AGN lightcurves are calibrated fall closely on these relations; \citealt{dimatteo:msigma,hopkins:bhfp.theory}), it is automatically implicit that they also reproduce the local BH mass function. An explicit calculation and comparison with the mass function estimated in \citet{marconi:bhmf} or \citet{shankar:bol.qlf} confirms this. This is also implicit since the extended ``continuity equation'' version of the \citealt{soltan82} argument \citep[see][]{yulu:bhmf,yutremaine:bhmf,merloni:synthesis.model} shows consistency between the quasar LF and BH mass function. Therefore this also has little power to constrain fueling models.

\vspace{-0.5cm}
\section{Discussion}
\label{sec:discuss}

%\vspace{-0.5cm}
\subsection{Overview}

This paper presents a simple ``semi-empirical'' population synthesis model for AGN fueling that distinguishes between major merger-triggered and non-major merger triggered (``stochastic'') activity. We show that this can plausibly account for the bolometric AGN luminosity function from $z=0-6$ and $\Lbol\sim10^{10}-10^{14}\,L_{\sun}$, observations of the distribution of AGN host morphologies, and observed AGN clustering amplitudes as a function of redshift and luminosity.

Our model builds on the ``semi-empirical'' model approach from \papertwo, which means that the ``background'' galaxy population properties are taken from observations. The theoretical ``layer'' added on top of this is the AGN fueling/feedback model. The non-major merger model is taken from \paperone; this model attempts to calculate the probability that cold, dense gas reaches an AGN and can be accreted, based on known empirical properties of galaxies (their distribution of gas fractions and the spatial distribution of that gas). The advantage of this model is that it makes no specific assumption about how this gas ``gets into'' the galaxy center in the first place -- it can be contributed or torqued by minor mergers, disk instabilities (bars, spiral arms, massive clumps), directly fueled by ``cold flows'' or accretion streams, or simply random turbulent cloud-cloud scattering. Since these all contribute in a {\em statistical} sense to the distribution of gas fractions and dispersions, they are all accounted for implicitly. The merger model is taken from \papertwo, using empirically-constrained merger rates convolved with a library of results from galaxy-galaxy merger simulations with simple prescriptions for BH growth and feedback.

%As noted, in \citet{kocevski:2012.candels.agn.hosts}, the original \paperone\ model under-predicts the luminosity below which disks dominate the AGN host population at $z\sim2$ by an order of magnitude. The updated model here, however, agrees well with (and even may slightly over-predict) the observed transition luminosity. What has changed? Recall, we have not altered the fundamental model itself (the calculation of the AGN duty cycle or lightcurves for fixed galaxy properties). The differences owe to better empirical inputs (namely the galaxy mass functions and gas fractions) into the models, and revised theoretical estimates of the merger rates. The largest difference is the gas fraction and mass function evolution: in \paperone\ the model assumed gas fractions were essentially constant at fixed stellar mass at higher redshifts, whereas in fact they appear to evolve strongly, which allows for considerably higher duty cycles of stochastic accretion. As well, the mass function does not appear to decline as steeply with redshift as had been assumed therein.

We reach similar conclusions to those recently reached by \citet{draper:2012.bh.synthesis.secvsmgr}, using an independent BH population synthesis approach with very different methods used to model the merger and non-merger triggering rates. In short, the models predict that ``stochastic'' fueling, with no specific preference for large-scale ``triggering phenomena'' in disky, secularly evolving systems should dominate the population at Seyfert and lower luminosities, while mergers dominate fueling of bright quasars. As argued in \citet{bellovary:2013.cosmo.bh.fueling}, this means that no new ``direct'' large-scale fueling mechanisms (such as cold flows somehow penetrating directly to the BH) need to exist at high redshift -- and in fact, there is little room for such mechanisms in this model. And recent observations in e.g.\ \citet{treister:2012} (and references therein) may have begun to map this transition -- the authors there see a strikingly similar trend with luminosity to that predicted here, with only a weak secondary dependence on redshift.

\vspace{-0.5cm}
\subsection{The Role of Stochastic Fueling as a Function of Mass/Luminosity}
\label{sec:discussion:stochastic.vs.mass}

Quantitatively, if we integrate the models here, we estimate that non-major merger AGN contribute about $\sim30\%$ of the total AGN luminosity density and BH mass density of the Universe. This agrees well with some recent observational estimates \citep{georgakakis:xr.agn.host.morph,koss:2010.swift.dual.agn}. But the predicted contribution of mergers is strongly BH mass and luminosity-dependent. Predicted low-mass BH growth is strongly dominated by non-major merger mechanisms, with nearly all the BH mass at $<10^{6}\,\msun$ and most of the BH mass at $<10^{7}\,\msun$ (at all redshifts) accreted in the ``stochastic'' mode. But above $M_{\rm BH}\gtrsim 10^{7}\,\msun$, most of the mass is accreted in the merger-induced mode. As argued in \S~\ref{sec:intro}, this seems physically reasonable. Growing a BH significantly above $\sim10^{8}\,\msun$ requires inflows that can channel a large fraction of an entire galaxy gas supply to $\lesssim10\,$pc in a Salpeter time -- essentially a single galaxy dynamical time! Galaxy interactions represent one of the only well-established and sufficiently violent mechanisms to accomplish this.

\vspace{-0.5cm}
\subsection{How Does This Relate to Star Formation?}

This may mirror a predicted and increasingly observationally well-established distinction in what powers galactic star formation. At low star formation rates, ``quiescent'' star formation (steady consumption of gas in disks) dominates, but the highest star formation rate systems are essentially all major mergers. At low-redshifts, this has been well-known for $\sim20$ years \citep[with the transition occurring at IR luminosities of ULIRGs, see e.g.][]{joseph85,sanders96:ulirgs.mergers,evans:ulirgs.are.mergers}. Models predict that the same should be true at high redshifts, but with a higher ``transition'' luminosity since all systems -- mergers and quiescent galaxies -- shift up to higher star formation rates at higher redshifts as all galaxies become more gas-rich (see e.g.\ \papertwo, and references therein). Observations have now progressively mapped this transition from $z\sim0-2$ \citep[see e.g.][]{tacconi:smg.mgr.lifetime.to.quiescent,younger:sma.hylirg.obs,casey:highz.ulirg.pops,melbourne:2008.dog.morph.smooth,dasyra:highz.ulirg.imaging.not.major,sargent:2012.ir.transition.lum.evol,zamojski:2011.highz.ir.gal.mgr.prop,kartaltepe:2012.high.disturbed.frac.highz.ulirgs}. 

However, there are two critical differences between the star-forming and AGN populations. First, the ``transition luminosity'' $L_{\rm SF}$ for star-forming populations (between ``quiescent'' star formation and merger-induced bursts) increases rapidly with redshift, rising from $\sim10^{11.5}\,L_{\sun}$ at $z=0$ to $\sim10^{13}\,L_{\sun}$ at $z>2$ (see references above). The predicted evolution in the AGN transition luminosity $\Lbol$ is much weaker (nearly constant at $10^{12}\,L_{\sun}$, in the model here). The rapid evolution in $L_{\rm SF}$ is widely attributed to the fact that, as gas fractions systematically increase at high redshift (itself owing to more rapid cosmological gas inflow rates), the associated star formation rates rise super-linearly according to the \citep{kennicutt98} relation.\footnote{This is the dominant effect driving evolution in the \papertwo\ models for the IR luminosity functions of star-forming galaxies, which appear to accurately describe the evolution in $L_{\rm SF}$.}
However, in most models, the maximum AGN $\Lbol$ is fundamentally limited by the BH mass (via the Eddington limit), {\em not} the galactic gas supply. Increasing gas fractions at high redshifts therefore tends to increase the AGN {\em duty cycle} in most models, but has relatively little effect on the characteristic {\em luminosities} of AGN \citep[see e.g.][]{hopkins:bhfp.theory,johansson:bh.scalings.in.remergers}. Since the mass at the ``break'' in the galaxy stellar mass function (hence implied BH masses, if the BH-host correlations still apply) does not evolve very strongly from $z\sim0-2$, this implies that the the AGN ``transition'' $\Lbol$ should be more constant than the star formation transition $L_{\rm SF}$.

Second, it is increasingly clear in both models and observations that the integrated total of star formation in the Universe is dominated by the ``quiescent'' mode. However, the integrated BH growth (at least in the model here) is dominated by the merger-induced mode. In the model, this is closely related to the origin of galactic bulges. Most of the total stellar mass in bulges is in ``classical'' bulges, which a wide range of observational and theoretical constraints indicate formed in violent mergers (see references in \S~\ref{sec:discussion:stochastic.vs.mass} above; for reviews, see \citealt{kormendy.kennicutt:pseudobulge.review,kormendy:2012.spheroidals,fisher:pseudobulge.ns,hopkins:seyfert.limits,balcells:bulge.scaling,gadotti:sdss.pseudobulge.properties}). However, even if most of the bulge is formed in such an event, it is primarily via the transformation of {\em pre-existing} stars from a disk to a bulge via violent relaxation. A wide variety of independent observations (including e.g.\ stellar age and metallicity distributions, kinematics, phase-space density profiles, gas density and star formation properties in ongoing mergers, and more) indicate that only a small fraction ($\sim10\%$ in an $\sim L_{\ast}$ spheroid) of the final stellar mass is actually formed in a nuclear starburst ``driven by'' the merger \citep[for a rigorous discussion, see][]{hernquist:phasespace,hopkins:sb.ir.lfs}. However, these inflows can dominate the formation of stars at extremely high densities in galaxy nuclei (much larger than the densities at the center of disks). And since it is {\em nuclear} inflows that ultimately matter for BH growth, these same inflows may dominate the growth of the BH population. 

Empirically, {\em if} it is true that BH mass is correlated with {\em bulge} mass (at the masses $\gtrsim10^{7}$ that contain most of the mass density), then it follows that most of the BH mass growth follows the mechanisms that build up most bulge mass (not necessarily the mechanisms that initially form those stars, if they are in disks). And most bulge mass is in classical (presumably merger-built) bulges. Though a subtle distinction, there is evidence that BH growth in luminous AGN is not strictly contemporaneous with most of the star formation, though they follow the same mean trends in a sufficiently time-averaged sense (as they must, for any linear BH-host mass relation); the sense is such that BH growth is biased towards more spheroid-dominated, and at high luminosities more obviously merging systems \citep[see e.g.][]{zheng:sfr.bhar.coevol.vs.coincidence,kartaltepe:agn.frac.vs.lir,santini:2012.agn.sf.corr.break}. In other words, this would say that most of the star formation is in low-mass, relatively low-luminosity galaxies, whereas most of the BH mass is in high-mass, bulge-dominated galaxies. However, it remains a critical, ultimately empirical question, to test whether BHs really do correlate with bulge (and not disk) properties, especially at higher redshifts \citep[for a recent review, see][]{kormendy:2013.mbh.mgal.review}.

\vspace{-0.5cm}
\subsection{Observational Predictions}

We have compared some of the lowest-order predictions of this model to observations, and reach a couple of important conclusions. First, the luminosity function itself, and the clustering (or environments) of AGN are {\em not} sensitive probes of the AGN fueling mechanism \citep[see also][]{bonoli:modeling.qso.clustering.vs.lifetimes.model}. Even when our model predicts a sharp transition in the fueling mode as a function of luminosity, no signature appears in these data. But we do predict a strong trend of the ``post-merger'' (bulge-like) versus disk population of AGN hosts as a function of luminosity, as discussed above. This is primarily a function of luminosity, and only weakly depends on redshift. At the moment, the statistical evidence for this is tantalizing, but not strong -- larger samples of systems with reliable morphological classifications, especially at high luminosities and redshifts, would be tremendously useful.

If this trend of bulge vs.\ disk-like AGN host morphologies is, in fact, due to the role of merger-induced fueling as we have proposed, there should be a number of corrollaries. The next step would be to examine the bulge-dominated hosts and look for evidence that they are not simply uniformly sub-sampling the ``normal'' bulge-dominated galaxy population. Indeed, there are already a number of additional, indirect observational suggestions that there is a transition from essentially random fueling of AGN at Seyfert luminosities to merger-induced fueling in true quasars; some of these are summarized in \citet{hopkins:seyfert.limits}. This includes the fact that quasars exhibit excessive small-scale (sub-halo scale) clustering while Seyferts do not \citep{serber:qso.small.scale.env,myers:clustering.smallscale,serber:qso.small.scale.env,hennawi:2009.new.binary.qsos,shen:2009.smallscale.qso.clustering.highz}; future observations should confirm that this difference appears even considering bulge-dominated galaxies. Quasar duty cycles rise more sharply with redshift (in agreement with observed merger rates), as opposed to Seyfert duty cycles which increase more slowly more or less in agreement with galaxy gas fraction evolution (see the compilation in \citealt{hopkins:seyfert.limits} and discussion in \citealt{draper:2012.bh.synthesis.secvsmgr}); this is qualitatively {\em opposite} the trend in galaxy bulge-to-disk ratios (galaxies become less bulge-dominated, even at fixed mass, at high redshift) -- it is therefore very difficult to predict a trend purely from secular or stochastic fueling in which the ratio of disk-to-bulge dominance is nearly redshift-independent (as we predict here). We also expect a much larger prevalence of ``post-starburst'' (or recently star-forming, K+A or E+A type) populations in true quasars compared specifically to normal {\em bulges} (not disks) of the same mass, as many observations have suggested \citep{brotherton99:postsb.qso,vandenberk:qso.spectral.decomposition,lutz:qso.host.sf,wang:highz.qso.ir,shi:qso.host.sf.lf,kewley:agn.host.sf,nandra:qso.host.colors,silverman:qso.hosts,higdon:postsb.qso,glikman:2012.red.qso.mergers,cales:2013.post.starburst.qsos}. 

The most naively obvious prediction to search for in the observations of these systems is evidence of tidal interactions, double nuclei, and other morphological merger signatures; in future work, we will make specific predictions for the prevalence of these signatures from the simulations used to inform the models here. However, as we cautioned in \S~\ref{sec:intro}, most models predict the quasars appear in the late stages of the merger (or even a few hundred million years after the coalescence of the galactic nuclei), since it takes many dynamical times for material to make its way to the center of the galaxy. And indeed, recent observations of local mergers in different stages have confirmed that the incidence of AGN is highly biased towards the post-merger phases \citep[during which almost all commonly-used morphological classifiers at high redshift would identify the galaxies as ``normal''; see e.g.][and references therein]{satyapal:2014.qso.vs.pair.separation}. This is because the timescale for tidal features, asymmetries, and other features to relax out to beyond the galaxy effective radius is relatively short; thus even most quasars triggered by major mergers are predicted to only have incredibly faint tidal features. For example, there are number of known nearby systems which are $<0.5\,$Gyr post-merger and have tidal features of surface brightness $\mu \gtrsim 28-30\,{\rm mag\,arcsec^{-2}}$ at $z=0$ \citep[see][]{schweizer98,schweizer:ngc34.disk}; \citet{bennert:qso.hosts} and \citet{canalizo:2013.recent.pops.in.merger.qsos} have identified a number of analogous cases among the nearest quasars (and indeed all of these cases were initially mis-classified based on early HST images as ``relaxed'' bulges). So it is almost impossible, unfortunately, for present morphological observations of bulges at $z>0$ to rule {\em out} late-stage or post-merger fueling. 

On the other hand, a particularly compelling corrollary of our model is the fact that, below the minimum BH mass required (by the Eddington limit) to power a quasar ($\sim 3\times10^{7}\,\msun$), most BH host bulges are ``pseudobulges,'' generally believed to form via secular processes (or minor mergers); above this mass, essentially all the bulges are ``classical,'' and so formed (at least initially) in (major) mergers \citep[see e.g.][]{kormendy.kennicutt:pseudobulge.review,kormendy:2012.spheroidals,fisher:pseudobulge.sf.profile,fisher:pseudobulge.ns,hopkins:seyfert.limits,
balcells:bulge.scaling,gadotti:sdss.pseudobulge.properties}. Verifying that this holds for the AGN hosts themselves (rather than simply for their ``relics'' at $z=0$, would represent a direct and very powerful confirmation of the models. Thus far this has only been confirmed for the very most local of samples, the PG quasars \citep[see][]{dunlop:qso.hosts,floyd:qso.hosts}. If bright quasar hosts at high redshift were instead all pseudobulges, a new form of fueling beyond what we model here may be required. 

\vspace{-0.7cm}
\acknowledgments 
Support for PFH was provided by NASA through Einstein Postdoctoral Fellowship Award Number PF1-120083 issued by the Chandra X-ray Observatory Center, which is operated by the Smithsonian Astrophysical Observatory for and on behalf of the NASA under contract NAS8-03060. KB was supported by the World Premier International Research Center Initiative (WPI Initiative), MEXT, Japan.\\

\vspace{-0.2cm}
\bibliography{/Users/phopkins/Documents/work/papers/ms}

\end{document}